# Frequency Comb-Based Remote Sensing of Greenhouse Gases over Kilometer Air Paths

G. B. Rieker[1,3], F. R. Giorgetta[1], W. C. Swann[1], J. Kofler[2], A. M. Zolot[1], L. C. Sinclair[1], E. Baumann[1], C. Cromer[1], G. Petron[2], C. Sweeney[2], P. P. Tans[2], I. Coddington[1], N. R. Newbury[1]

[1]*National Institute of Standards and Technology, Boulder, CO*
[2]*National Oceanic and Atmospheric Administration, Boulder, CO*
[3]*University of Colorado, Boulder, CO*

**Abstract**
We demonstrate coherent dual frequency-comb spectroscopy for detecting variations in greenhouse gases. High signal-to-noise spectra are acquired spanning 5990 to 6260 cm$^{-1}$ (1600 to 1670 nm) covering ~700 absorption features from $CO_2$, $CH_4$, $H_2O$, HDO, and $^{13}CO_2$, across a 2-km open-air path. The transmission of each frequency comb tooth is resolved, leading to spectra with <1 kHz frequency accuracy, no instrument lineshape, and a 0.0033-cm$^{-1}$ point spacing. The fitted path-averaged concentrations and temperature yield dry-air mole fractions. These are compared with a point sensor under well-mixed conditions to evaluate current absorption models for real atmospheres. In heterogeneous conditions, time-resolved data demonstrate tracking of strong variations in mole fractions. A precision of <1 ppm for $CO_2$ and <3 ppb for $CH_4$ is achieved in 5 minutes in this initial demonstration. Future portable systems could support regional emissions monitoring and validation of the spectral databases critical to global satellite-based trace gas monitoring.

**Introduction**

Absorption spectroscopy over open paths provides a means of remotely sensing changes in greenhouse gas mole fractions – a critical need for climate change science[1]. It is implemented in satellite instruments, upward-looking Fourier Transform Spectrometers (FTS), and ground-based FTS and laser spectrometers[2–9]. Ideally, these systems should detect the dry-air mole fractions (which correct for variable dilution by water vapor) of multiple gases over long paths with high precision and reproducibility to enable mapping of small gradients in both space and time. However, absorption spectroscopy faces two distinct challenges. First, spectral databases required to convert absorption to concentrations cannot always support the desired reproducibility between instruments, for example below 1 ppm for $CO_2$[10–13]. Second, there are no portable, long-path, multi-gas sensors with high reproducibility. Portable FTS is limited to sub-km paths and ~5-10% uncertainty because of divergent sources and broad instrument lineshapes[8,14]. Therefore, regional studies use flushed-cell point sensors calibrated via a reference gas[15,16]. Accurate open-air path systems could provide continuous path-averaged mole fractions that avoid representation errors associated with point sensors[1,17].

Dual frequency-comb spectroscopy (DCS) is a promising solution to both challenges. DCS[18–29] has broadband spectral coverage for multi-species detection, a bright diffraction-limited source for high signal-to-noise ratio over multi-kilometer ranges, a rapid update rate for immunity to turbulence-induced optical intensity fluctuations, and, importantly, can sample the transmission on a comb tooth-by-tooth basis for high-accuracy spectra. Here, we show that the full advantages of DCS can be applied to quantitative outdoor sensing of greenhouse gases. Our measured spectra span 80,000 comb teeth covering 5990 to 6260 cm$^{-1}$ (1600 to 1670 nm) with absorbance noise below 5×10$^{-4}$. Data are acquired at the comb tooth separation of 0.0033 cm$^{-1}$ (100 MHz), with negligible instrument lineshape since the comb teeth are essentially delta-functions in frequency. We demonstrate simultaneous retrieval of dry-air mole fractions of $CO_2$, $CH_4$, $H_2O$, HDO, $^{13}CO_2$ and air temperature over a 2-km turbulent air path. During well-mixed atmospheric conditions, these data enable high-resolution, broadband evaluation of spectral absorption models and, when combined with laboratory measurements, should lead to improved spectral absorption models critical for open long-path remote sensing[11,30].

Moreover, the advent of portable frequency combs[31] should enable field-deployable DCS to support verification and monitoring of emissions of distributed sources (e.g. carbon sequestration[9] and gas development sites[32,33]) and larger-scale monitoring networks. As an initial demonstration, time-resolved dry-air mole fractions were acquired continuously over three days. The DCS data compares well with a nearby point sensor for large-scale fluctuations with much lower sensitivity to local concentration spikes. One-sigma stabilities of <1 ppm (µmol/mol) for $CO_2$ and <3 ppb (nmol/mol) for $CH_4$ are reached at five-minute averaging. Absolute agreement is limited by the current spectral databases to ~1-2% and by variability in sampled regions. Future optimized systems with higher power and extended spectral coverage[27,34–39] could reach similar stabilities in seconds, over 10 kilometers, while sensing additional species, isotopologues, and oxygen[26]. Finally, the absence of instrument lineshapes should enable direct cross-comparison of retrievals between systems, times, and locations.

**Open-Air Dual Comb Spectrometer**

Figure 1 shows the experimental setup. In dual-comb spectroscopy[18–28], two frequency combs are arranged to have offset repetition rates ($f_r$ and $f_r+\Delta f_r$). When combined, the resulting heterodyne signal is an rf frequency comb, where each rf comb tooth has a one-to-one relationship with a specific pair of optical comb teeth (see Fig. 1a). Therefore, this rf spectrum is simply scaled to reproduce the optical spectrum. Here, we implement DCS with two mutually coherent Erbium-doped fiber frequency-combs ($f_r$~100 MHz, $\Delta f_r$ = 444 Hz) with relative ~ 1-Hz comb linewidth and with each comb tooth's frequency known to better than 1 kHz[21] (see Appendix). The comb spectra are centered within the atmospheric water-vapor window near 1.6 µm and further shaped to cover both a portion of the $CH_4$ tetradecad and a $CO_2$ combination band. The comb outputs are combined and transmitted over a 2-km open air path on the NIST Boulder campus (see Figs. 1b-c). This 2-km pathlength exceeds previous laboratory-based DCS using either multi-pass cells or resonant cavities.

Figure 1d-e shows the resulting high signal-to-noise ratio (SNR) transmission spectrum acquired over ~170 minutes under relatively constant temperature and pressure across the measurement path (see Appendix). The overall shape corresponds to the spectrally filtered comb light. The stronger $CO_2$, $H_2O$ and $CH_4$ absorption lines appear as sharp dips of up to 15% with many weaker $CH_4$, $CO_2$, and $H_2O$, HDO and $^{13}CO_2$ lines observed down to $10^{-3}$ absorbance. There are ~80,000 comb tooth pairs across the 267 cm$^{-1}$ window. With the spectral shaping, about half of these, or ~40,000 comb tooth pairs compose the measured spectrum. Each comb-tooth pair contributes a distinct data point in the transmitted spectrum with kHz-level frequency uncertainty (corresponding to a resolving power of $10^{11}$), spaced at $f_r$ =100 MHz (0.0033 cm$^{-1}$), and with SNR in the signal intensity exceeding 3000:1 for these long time-averaged data. Even higher SNR values would be possible except that we aggressively limited the transmitted power to avoid any lineshape distortions due to detector nonlinearity[21,27]. These data were acquired with coherent summing (see Appendix), but continuous time data confirmed ~Hz-linewidth between the detected pairs of frequency comb teeth. The quality of these data is consistent with previous ultra-high resolution laboratory DCS spectra and demonstrates the fundamental properties of coherent DCS –

namely high resolution, high accuracy, broad bandwidth, and high SNR – can be directly translated to field-based measurements.

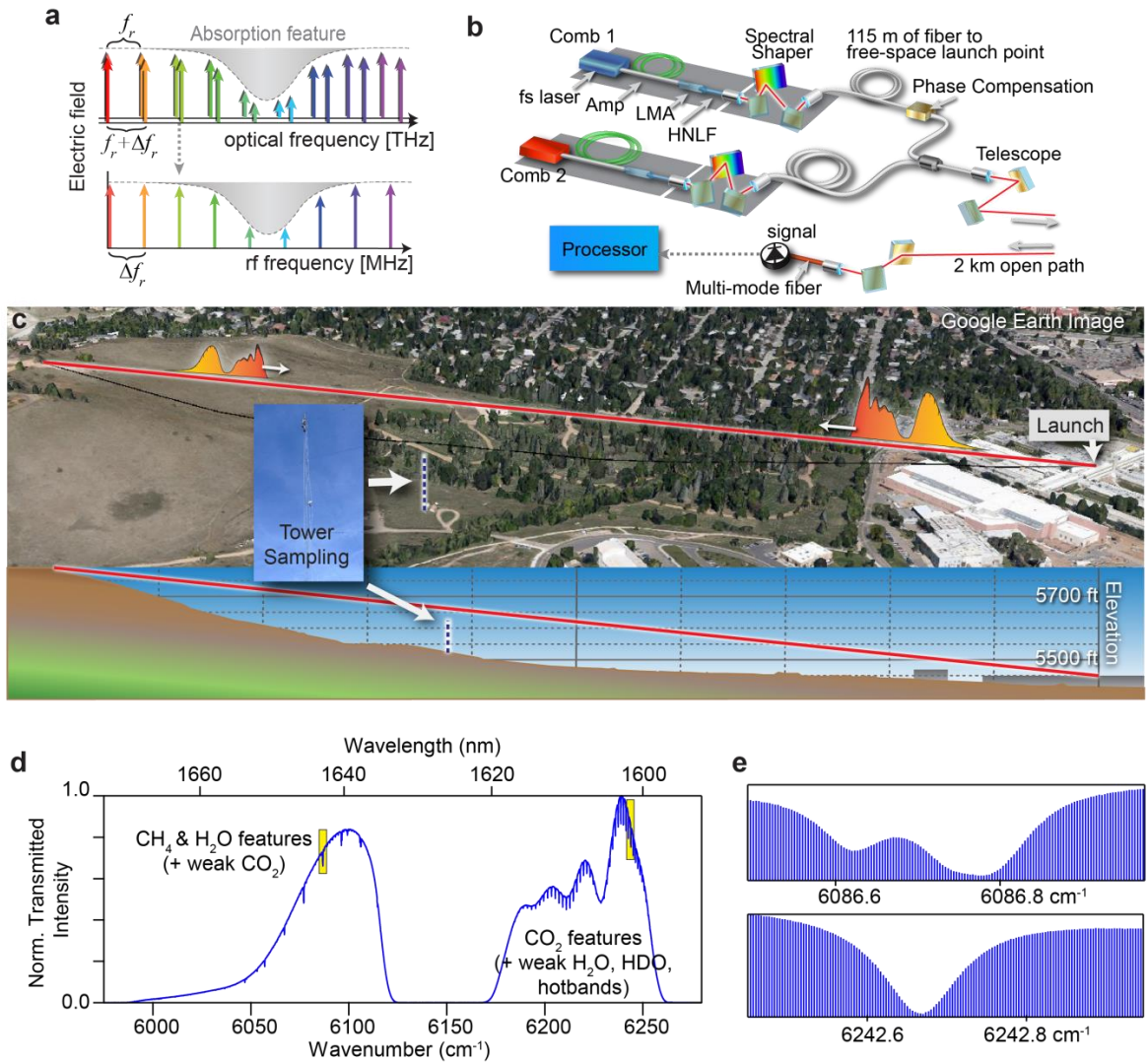

Figure 1. Open air-path greenhouse gas sensing through dual comb spectroscopy. a. DCS concept: two combs with slightly different tooth spacing interfere on a detector, giving a third rf comb with a one-to-one mapping to the optical comb teeth. (The actual experiment spans ~$10^5$ comb teeth.) b. Experimental setup: two combs are amplified, pulse-compressed in large mode area (LMA) fiber, spectrally broadened in highly nonlinear fiber (HNLF) and filtered to generate light covering the spectral bands of interest. Two 115 m fibers carry the comb light to the rooftop where the light is combined and launched in a ~40-mm beam ($1/e^2$ diameter). The return light is coupled to a 62-μm graded-index multimode fiber and detected. The transmitted light power was limited to 1.5 mW so that the maximum received power was always below a conservative detector nonlinearity threshold of 50-μW average power. c. The location of the 2-km interrogation path (red line, ground projection represented by black line), the tower with the point sensor intake (inset), and elevation of the beam path (bottom inset). d. Example transmitted intensity showing the smoothly varying comb intensity and abrupt dips due to absorption. e. Expanded view across two absorption features. The typical gas absorption lines have ~ 0.13-cm$^{-1}$ width compared to the 0.0033-cm$^{-1}$ comb tooth point spacing yielding ~40 teeth across each line (top: several transitions from the $2\nu_3$ level of the $CH_4$ tetradecad, bottom: R20 transition of the 30013←00001 band of $CO_2$).

Turbulence is a concern for high-resolution open-air path spectroscopy as it can easily cause strong (100%) and fast (>100 Hz) optical intensity modulation, potentially leading to excess noise in the measured optical transmission spectrum. The power spectral density related to turbulence-induced intensity noise falls off strongly as $f^{-8/3}$ beyond the characteristic cutoff Fourier frequency, $f_c \approx U/\sqrt{2\pi L\lambda}$, where $U$ is the wind speed, $\lambda$ is the optical wavelength, $L$ is the pathlength[40]. Here, $f_c$ is tens of Hz. In comparison, the DCS effectively acquires a single spectrum within $1/\Delta f_r = 1/444$ s. In other words, since $f_c < \Delta f_r$, the turbulence intensity noise is effectively frozen during a single spectrum. Of course, some turbulence-induced fluctuations do occur on the timescale of a single interferogram, but a more rigorous discussion (see Fig. 5) shows they appear as multiplicative noise that is below the overall noise floor. Turbulence can also cause optical wavefront distortions and phase noise on the comb lines[40,41]. However, since both combs are co-propagating, these effects are common mode and ultimately negligible. This relative immunity of DCS to turbulence is in strong contrast to high-resolution FTS or conventional swept laser spectroscopy, which have longer acquisition periods.

**Measurements under well-mixed atmospheric conditions**

The DCS spectra provide a direct means to validate current and future spectral databases and absorption models since the spectra are free from instrument distortions and the DCS horizontal path avoids the atmospheric modeling that is required with up-looking total column measurements[5,12]. The transmission spectrum shown in Fig. 1d is converted to absorbance through piecewise baseline fitting. (See Appendix) The resulting absorbance spectrum, shown in Fig. 2, can then be fit using different absorption models to both assess those models and find the path-averaged dry-air mole fractions of various greenhouse gas species.

We separately fit the lower (6050 to 6120 cm$^{-1}$) and upper (6180 to 6260 cm$^{-1}$) spectral window. Within a spectral window, we fit the entire absorption spectrum at once. The only inputs to the fits are the measured atmospheric pressure (from calibrated pressure gauges at each end of the path) and the absorption models. The fitted parameters are the overall gas concentrations of $CO_2$, $^{13}CO_2$, $H_2O$, $CH_4$ and HDO. The fit to the upper spectral window also includes a fit for temperature based on the band-wide $CO_2$ absorption; in this way the path-averaged temperature is extracted directly instead of using co-located temperature sensors that can suffer from solar loading. The $CO_2$ and $^{13}CO_2$ mole fractions are extracted from the fit to the upper spectral region, while the $CH_4$, $H_2O$ and HDO mole fractions were extracted from the fit to the lower spectral region (although there are $CO_2$ lines present as well). More than 300 spectral lines are included in the lower spectral window and 400 in the upper spectral window.

Absorption models for species in this region are evolving. The models consist of a set of spectral parameters that describe the temperature-, pressure-, and concentration-dependent strength, location, and width of each absorption feature, along with a lineshape profile model. Figure 2 shows the results of fits using several different absorption models. The standard deviation of the residuals is ~$2\times10^{-3}$ absorbance units for 5-minute averages at the 100-MHz (.0033 cm$^{-1}$) point spacing, dropping to <$5\times10^{-4}$ at 170 minutes. When scaled to the same resolution, this SNR is comparable to high-resolution solar, up-looking FTS spectra (but in a more compact, potentially portable instrument package without long, moving interferometer arms). For the upper spectral region, the very similar absorption models based on Hitran 2012[42] and Hitran 2008[43] result in residuals with similar peak values below $3\times10^{-3}$, except for one errant line near 187.38 THz (coincident with a reported weak HDO line in Hitran 2008). For the lower spectral region, Hitran 2012 has methane parameters quite different from Hitran 2008, and this difference is strongly reflected in the residuals, as well as the concentrations as discussed in Table I.

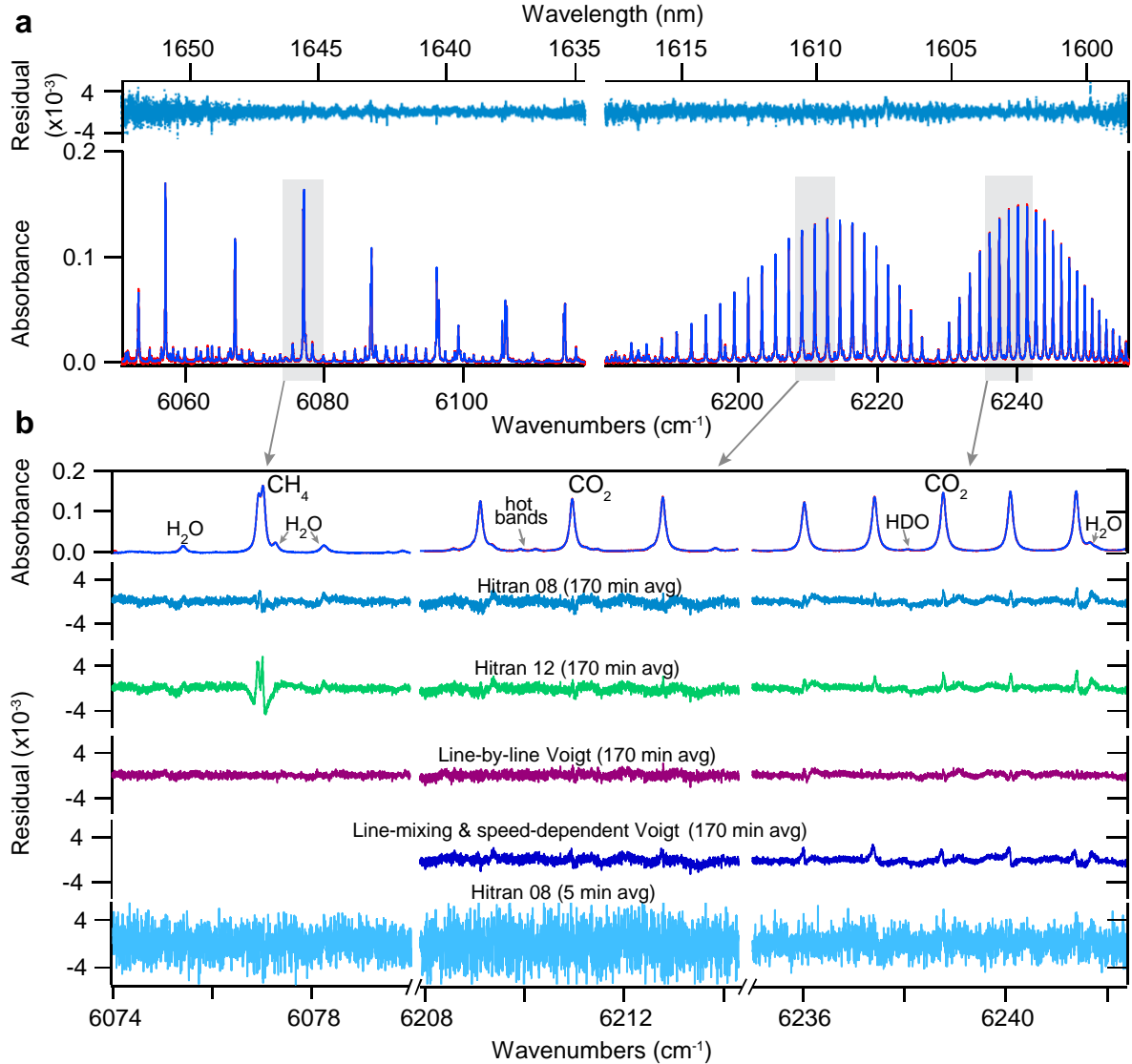

Figure 2: Baseline-corrected absorbance from the transmission spectrum of Fig. 1d. **a.** The left spectrum mainly comprises $CH_4$ and $H_2O$ lines, with some weaker $CO_2$ lines, while the right spectrum shows the 30013←00001 $CO_2$ band, $CO_2$ hot bands, $H_2O$, $^{13}CO_2$ and HDO features. The data (red line), fit with Hitran 2008 (blue line, indistinguishable from the data), and residuals (upper blue line) are shown. There are ~40,000 data points, or comb teeth, spaced at 100 MHz across the spectral windows. **b.** Expanded view of a few representative lines along with residuals for fits to the HITRAN 2008 model[43], HITRAN 2012 model[42], line mixing and speed-dependent model of Ref.[12], and line-by-line Voigt fits to each line. Also shown are the fit residual for a 5-minute time average, identical to that used in the time-resolved data of Fig. 3. The noise per comb tooth (in absorbance units) for the 5-minute averaged spectra is $2.0 \times 10^{-3}$, $2.7 \times 10^{-3}$, and $1.7 \times 10^{-3}$ for the three windows shown, with the differences attributable to different spectral intensities. For the 170-minute average, the noise decreases with the square root of time, as expected, to $3.5 \times 10^{-4}$, $5.1 \times 10^{-4}$, and $3.0 \times 10^{-4}$, respectively. However, there are clear residuals near the spectral lines due to mismatch between the measured lineshapes and absorption models.

The standard Hitran models[42,43] for the measured bands neglect, among other things, coupling between energy states of nearby transitions (line mixing) and the effect of collisions on the Doppler contribution to the lineshape (speed dependence). Therefore, concentrations extracted using these databases with Voigt profiles have been shown to lead to inaccurate atmospheric retrievals[44]. Figure 2b also shows residuals to a fit using a lineshape model and corresponding spectral parameter database that includes line mixing and speed-dependent Voigt profiles and was developed to support accurate satellite-based trace gas monitoring[12,45,46]. Residuals remain, with quite different structures compared to the other models, however the fitted concentration using the line mixing and speed dependent model should have the highest accuracy.

Figure 2b also contains residuals resulting from a line-by-line Voigt profile fit where collisional linewidth, line center, and line strength are finally allowed to vary on a line-by-line basis. Though this fit overlooks the quantum-mechanical basis of the previous spectral databases, the small residual may indicate that the largest sources of error in absorption models are the spectral parameters (line strength, line center, broadening coefficients, etc.) rather than deviations from the Voigt lineshape model.

Table I compares the mole fractions extracted using the different absorption models. There is a significant model-dependent spread. For $CO_2$, all four absorption models rely on analysis of the same underlying laboratory FTS data, so this spread emphasizes the consequences of different lineshape models. As mentioned above, among the models considered here the line-mixing, speed-dependent Voigt profile (LM/SD) and corresponding spectral parameter database[12,45,46] is expected to yield the most accurate results for $CO_2$. Table I also reports the DCS systematic uncertainty excluding the model-dependent effects. This uncertainty is based on the sensitivity of the retrieved mole fractions to maximum pressure and temperature path inhomogeneities (±500 Pa and 6 K, respectively), uncertainty in path length (±20 cm) and air pressure (±30 Pa), and baseline correction. Baseline correction is the largest contributor (by a factor of 10 or more) in particular due to an etalon ripple with absorbance amplitude $10^{-3}$ (see Appendix); For $CO_2$ the other factors contribute below 0.06 ppm uncertainty.

|  | Mole Fraction Retrieval (ppm) | | | | Systematic Unc. (ppm) | |
| --- | --- | --- | --- | --- | --- | --- |
|  | Hitran 08 | Hitran 12 | LM / SD | Toth | excluding spectral model | |
| $CO_2$ | 408.7 | 407.7 | 404.7 | 406.2 | 0.8 | 0.21% |
| $CH_4$ | 1.878 | 1.985 | -- | -- | 0.009 | 0.45% |
| $H_2O$ | 3223 | 3217 | -- | -- | 22 | 0.73% |
| HDO | 1.13 | 0.97 | -- | -- | 0.13 | 11% |
| $^{13}CO_2$ | 4.5 | 4.4 | -- | 4.2 | 1.7 | 37% |

Table I: Mole fractions retrieved from fits of the absorbance spectrum of Fig. 2 with four absorption models: HITRAN 2008[43], HITRAN 2012[42], line-mixing speed-dependent Voigt (LM/SD)[12,45,46] and Toth et al.[47]. The final column reports the systematic (type B) uncertainty of the dual-comb spectrometer, excluding the model dependence captured in the first columns (see text).

Table II compares the DCS to the tower-mounted point sensor. For $CO_2$, there is 1.8% offset with the LM/SD fits, which increases to 2.8% for the Hitran 2008 fits. Under the windy, well-mixed atmospheric conditions for these data, the DCS and tower-mounted point sensor should measure almost identical mole fractions. Given the point sensor is calibrated directly against the World Meteorological Organization (WMO) reference gas (see Appendix), we attribute most of the offset to the DCS retrieval and specifically to the line strengths of the absorption model. For $CH_4$, the DCS analyzed with the Hitran 2008 database is in excellent agreement with the point sensor (although the fits to Hitran 2012 exhibit a 5.7% offset from the tower sensor). For $H_2O$, the two systems agree to within the uncertainty of the tower sensor.

The main conclusion – that better absorption models are needed to support accurate greenhouse gas monitoring– very much echoes the significant body of work in support of satellite measurements. Sub-percent uncertainties in retrieved gas concentrations will require improvements in the spectral database, possibly through laboratory frequency comb or cavity ring down systems[21,27,48,49]. Finally, while DCS does rely on accurate spectral databases (as with any open-air path absorption technique), it should be straightforward to reanalyze DCS spectra as the databases are refined. In fact, this feature of the dual-comb spectra is important for accurate greenhouse gas monitoring. Whereas extractive flushed-cell sampling instruments rely on reference gas calibrations that must be performed periodically to maintain accuracy and cross-instrument comparability, the measured dual-comb spectra from multiple instruments can be compared directly and indefinitely. Therefore, the retrieved mole fractions can be similarly compared when the spectra are fit with an accurate, bias-free absorption model. For example, as water-broadening coefficients for $CO_2$ and $CH_4$ become available, these effects can be included without the empirical corrections needed with flushed-cell point sensors[16].

| Mole Fraction (ppm) | DCS | Tower Sensor | Difference | |
|---|---|---|---|---|
| $CO_2$ | 404.7 ± 0.8 | 397.6 ± 0.06 | 7.1 | 1.78% |
| $CH_4$ | 1.878 ± 0.009 | 1.874 ± 0.002 | 0.004 | 0.20% |
| $H_2O$ | 3223 ± 22 | 3168 ± 95 | 55 | 1.74% |

Table II: Comparison of the mole fractions obtained with the dual-comb spectrometer and the tower-mounted point sensor under the well-mixed, stable atmospheric conditions of Fig. 2. For the DCS, the dry-air mole fraction of $CO_2$ is retrieved from the LM/SD fit, and the dry-air $CH_4$ and $H_2O$ ratios are retrieved from the HITRAN 2008 fit.

**Time-resolved mole fractions over a three-day period**
The data of Fig. 2 were acquired over a windy, well-mixed period where the mole fractions were quite stable. Normally, the mole fractions will vary significantly from nearby sources and sinks, and as the planetary boundary layer height changes. We can analyze these time variations using the same fitting procedures described above. Figure 3 shows the results over a three-day period at five-minute averaging, analyzed with the HITRAN 2008 spectral database. The path-averaged air temperature is extracted directly from the fit to the 30013←00001 $CO_2$ spectral band, placing even greater reliance on accurate spectral parameters.

The tower-mounted point sensor records the mole fractions for $CO_2$, $CH_4$, and $H_2O$, which are over-plotted in Fig. 3. During periods of wind and daytime thermal turbulence, both the DCS and tower sensor mole fractions are relatively flat. During low wind and nighttime boundary-layer settling, both show strong variations in time corresponding to heterogeneous gas buildup and plumes. However, as expected, the point sensor is much more susceptible to short spikes from local plumes. (For comparison purposes, the tower-sensor data is averaged to 5 minutes here; at shorter timescales the spikes are even more pronounced.) Moreover, statistically significant offsets are common between the point sensor and the path-averaged DCS results, which is not surprising given the presence of localized emissions from vehicles, the NIST central utility plant (just south of the air path), and a nearby roadway. The comparison emphasizes the quantitative differences that can arise between a single point sensor and a path-averaged system. One expects the path-averaged system to connect more directly to km-scale regional transport models. Moreover, with the addition of multiple reflectors the same DCS system could interrogate multiple displaced paths[9], providing even greater utility to regional models.

The overall sensitivity, or precision, of the DCS is calculated directly as the Allan deviation over a period of relative stability, as shown in Fig. 4. The optimal averaging time period depends on the timescale of the atmospheric fluctuations; we selected 5 minutes for Fig. 3, which also leads to a precision comparable to the systematic uncertainty (Table II). However, operation over longer distances, at the full eye-safe power level of 9.6 mW, or at longer wavelengths where the absorption is tenfold stronger, will all dramatically improve the precision. In addition, stronger absorption lines will reduce the systematic uncertainty that is dominated by baseline ripple (etalons).

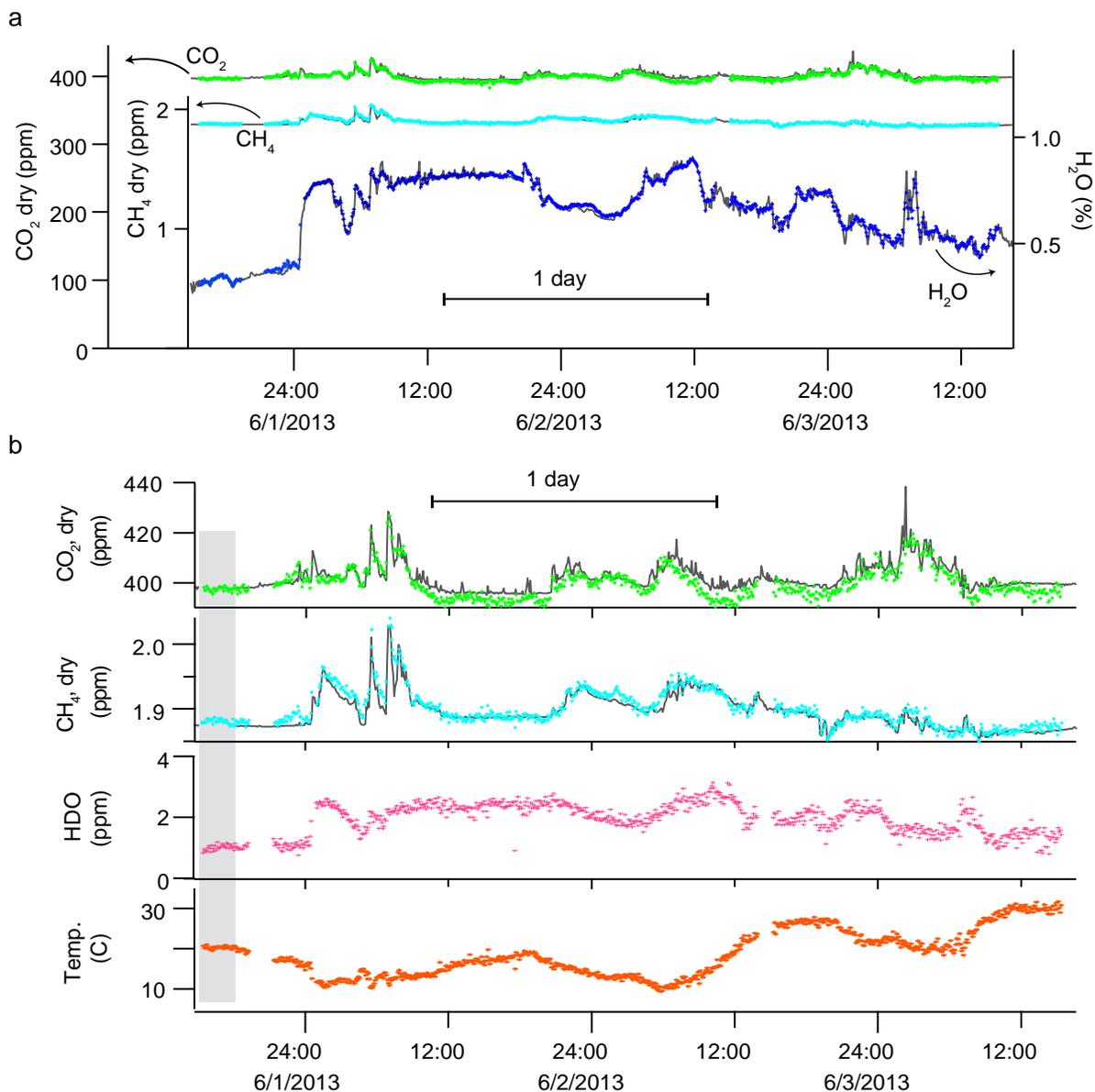

Figure 3. Time-resolved mole fractions. **a** Time dependence of the dry-air mole fraction for $CO_2$ (green), $CH_4$ (light blue) and water (dark blue) retrieved from the DCS and the corresponding values from the tower-mounted point sensor (black line) at 5 minute periods over three days. **b** Time-dependence of $CO_2$ and $CH_4$ with an expanded y-axis. Also shown are the time dependence of HDO (pink) and the retrieved path-averaged temperature (orange). All data are from the fit with the absorption model based on HITRAN 2008. The DCS $CO_2$ mole fraction has been scaled to remove the 2.8% offset between the DCS and tower-mounted sensor over the initial ~ 3-hour period (shaded region) where the atmosphere was well-mixed and the measurement path upwind from the NIST central utility plant.

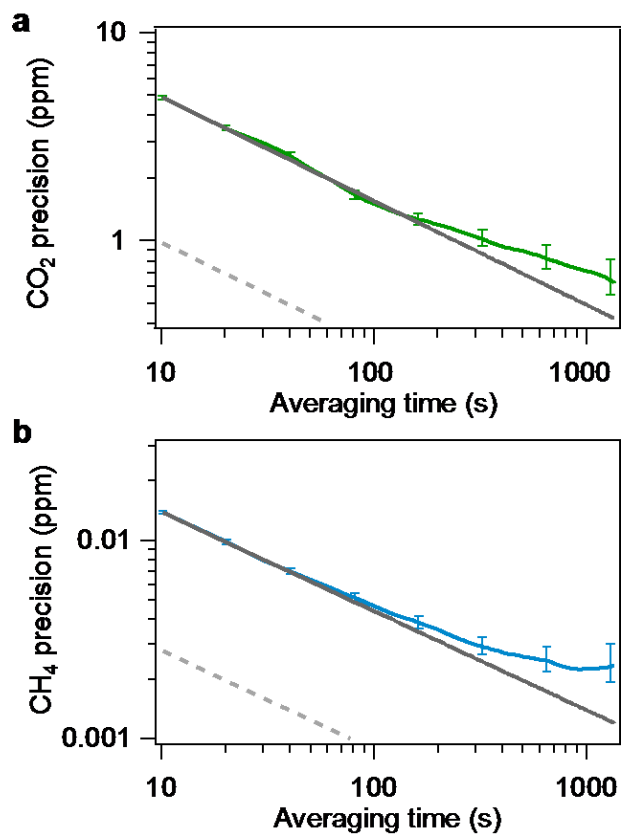

Figure 4. Precision (Allan deviation) from data acquired during high-wind periods with a stable, well-mixed atmosphere. **a** $CO_2$ precision (green) versus averaging time, $\tau$, which follows $15/\sqrt{\tau}$ ppm, where $\tau$ is in seconds (grey line) or 0.86 ppm at 5 minutes. At longer times, actual atmospheric mole fraction variations cause a decrease in the slope. A modest increase of the launched power up to the eye-safe limit of 9.6 mW (or alternatively an increased receive aperture) would yield the improved precision of $3/\sqrt{\tau}$ ppm (dashed grey line). **b** $CH_4$ precision (blue) follows a scaling of $40/\sqrt{\tau}$ ppb (grey line) or 2.3 ppb at 5 minutes. Again, modest power increases, or increased receive aperture, would yield the improved precision of $8/\sqrt{\tau}$ ppb (dashed grey line).

**Conclusion**

We demonstrate that dual frequency-comb spectroscopy is ideally suited to the challenges associated with accurate sensing of atmospheric trace gases across open-air paths; it combines broadband spectral coverage for accurate multi-species detection, with a diffraction-limited laser output for high sensitivity over multi-kilometer air paths. Equally important, its high acquisition speed achieves immunity to turbulence-induced intensity noise and its negligible instrument lineshape enables high accuracy and ultimately accurate cross-comparison of spectra (and therefore concentrations) acquired with different systems, at different times and locations. We demonstrate these capabilities over a 2-km open air path with an initial system that measures $CO_2$, $^{13}CO_2$, $CH_4$, $H_2O$, HDO and air temperature. Future broader bandwidth systems will detect more species, while higher power output will further improve the sensitivity. DCS data can support the development of accurate absorption models used in global, satellite-based greenhouse gas monitoring. Moreover, with the recent advancement of portable, high-performance frequency combs[31], there is no technological barrier to regional deployment of fielded DCS systems that have costs comparable to high-performance point sensors and which are capable of autonomous, eye-safe monitoring of multiple gas species over multiple optical paths.

**Acknowledgements**
G.B.R. received partial support from the National Research Council Research Associateship award. This work was funded in part by the NIST Greenhouse Gas and Climate Science Measurements Program. It benefited from technology developed under the DARPA PULSE program. We thank James Whetstone, Joe Hodges, and Scott Diddams for helpful discussions, Anna Karion for calibration of the point sensor, Terry Bullett for use of the radio tower, and Masaaki Hirano and Jeff Nicholson for donation of specialty optical fiber.

**Appendix**
**Effect of turbulence-induced intensity noise**

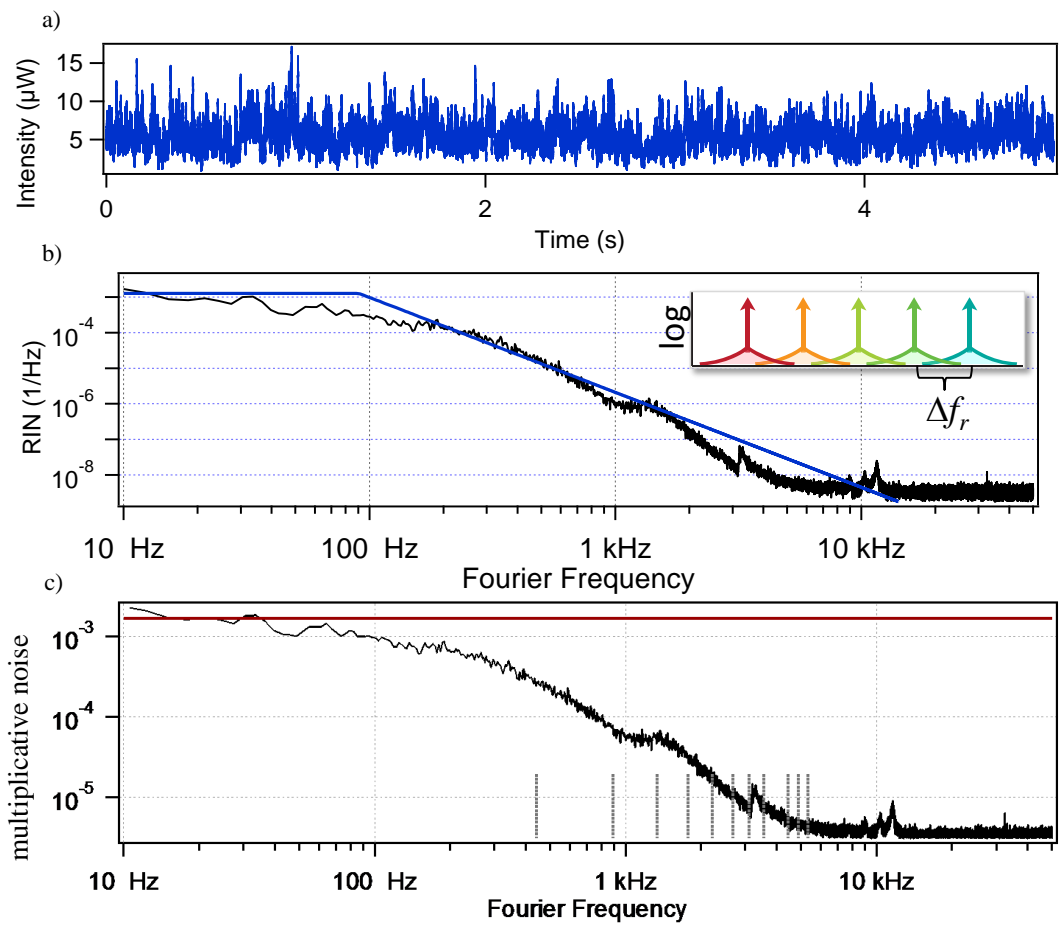

Figure 5: Effect of turbulence-induced intensity noise on dual-comb spectroscopy signals. **a.** Measured received intensity, $I$, across the path at 400 ksamples/second. **b.** Relative intensity noise, RIN, calculated as the power spectral density of $(\delta I)^2/\langle I\rangle^2$ (black line). Also shown is the theoretical power spectral density for a spherical wave at 5 m/s wind velocity, over a 2-km path, with a turbulence structure constant $C_n^2 = 10^{-15}$ m$^{-2/3}$ (blue trace)[40]. The corner frequency at $f_c = 2.5U/\sqrt{2\pi\lambda L}$ is clearly evident. This RIN (or its square root) acts as a strong amplitude modulation on the measured rf heterodyne comb teeth of Fig. 1a. The result is that some small fraction of the power of one rf comb tooth occurs at the adjacent rf comb tooth, leading to additional multiplicative noise from turbulence. (See inset.) **c.** Multiplicative noise between rf comb teeth separated by an rf frequency, $f$, for a 5-minute averaging

time calculated as $\sqrt{RIN(f)/300s}$ (black line). The actual rf comb tooth positions at integer multiples of $\Delta f_r =$ 444 Hz are illustrated as dashed gray lines. The multiplicative noise between adjacent comb rf comb teeth is $2.6\times10^{-4}$ and falls off sharply for distant comb teeth. This noise level is well below the contribution from other noise sources, equal to ~$1.7\times10^{-3}$ (from Fig. 2 and shown as a horizontal brown line). The ratio between the multiplicative noise and the measured noise floor is ~10× here, indicating the system is dominated by additive noise; therefore, only at 20 db stronger turbulence would the turbulence-induced multiplicative noise exceed other noise sources. Operation at a higher comb-tooth spacing (repetition rate) of 200 MHz would permit a 4× increase in $\Delta f_r$ and an additional order-of-magnitude immunity to turbulence noise.

**Experimental setup**

The frequency combs were stabilized by locking two teeth of each comb to two cavity-stabilized CW lasers at 1535 nm and 1560 nm, which provided both the mutual coherence between the combs and frequency accuracy[21,27]. Similar performance could be achieved through two self-referenced combs, each locked to a free-running cw laser to attain the necessary mutual coherence[24] but deriving their frequency accuracy from a stable rf oscillator. The combs were located in a laboratory and transmitted over two separate 115-m long fibers (to avoid cross-phase modulation nonlinearities) to the rooftop laboratory. Fiber-path fluctuations were monitored with a cw laser interferometer and cancelled with a PZT fiber stretcher to maintain the ~1-Hz relative comb linewidth. Both telescopes included a fast steering mirror to correct for building sway and slower turbulence-induced pointing effects. The X and Y axis of both fast steering mirrors were dithered between 55 and 95 Hz resulting in a few-percent dither on the received power, which was de-modulated to generate an error signal for control of the fast steering mirrors at a ~ 1 Hz bandwidth.

The data are digitized continuously at one comb's repetition rate. Phase-locking of the combs was arranged such that interferograms repeated phase-coherently exactly every $N$=225,000 points and could be coherently averaged point-by-point with a field-programmable gate array (FPGA) in real time for times up to the ~0.5 s mutual comb coherent period before saving[21,27]. A linear phase correction is applied to these 0.5 second-averaged interferograms before they are summed further with separate corrections applied for the $CO_2$ and $CH_4$ spectral windows because of differential delay from the 115-m optical fiber.

**Flushed-Cell Point sensor**

The point sensor was a commercial cavity-ringdown spectrometer. Its $CO_2$ and $CH_4$ response were calibrated prior to the measurements with four reference tanks of different mole fraction calibrated on the World Meteorological Organization scales at the National Oceanic and Atmospheric Administration. At the end of the measurements, the sensor was again tested against a single reference cylinder and the measured drift (which is larger than the uncertainty in the initial calibration) is reported here as the uncertainty. No direct calibration of water vapor measurement was performed but it is assumed from Ref.[50] to be ±3%.

**Spectral fitting**

The absorbance (Fig. 2) is calculated from the transmission spectrum (Fig. 1d) by taking the natural logarithm of the transmission spectrum and simultaneously fitting across 150-200 GHz windows with a 6$^{th}$-order polynomial baseline (including an absorption model calculated at the nominal conditions). This baseline correction is then applied to the entire spectrum to yield the absorbance profile. In an alternative approach, we acquired a separate parallel reference spectrum (not shown in Fig. 1b) for normalization, but this approach led to excess etalons and the aforementioned baseline fit was found to be superior. (Note that to avoid additional etalons in the data, the building windows were open during the measurement.) To determine the uncertainty associated with this baseline fitting, we processed the data in different ways. We modified the polynomial order, split the data into different subsets, adjusted the spectral window positions, and alternatively fit out or did not fit out the $10^{-3}$ absorbance ripple (etalon). Based on these analyses, we find an associated uncertainty of below 0.1% for the strong $CO_2$ and $CH_4$ lines but much higher uncertainty for the weaker lines from other gas species (see Table I).

To fit the spectra, a matrix of simulated spectra is generated using the selected database and spectral lineshape model (either Voigt or line-mixing and speed-dependent Voigt) for different air pressure, humidity, and temperature conditions at nominal pressure of the fitted species (e.g. $CO_2$, $CH_4$, $H_2O$, HDO, and $^{13}CO_2$). We then perform a least-squares fit to the upper spectral region, allowing the species mole fraction and temperature to float. (The species mole fraction floats via a linear scaling factor that neglects small changes in self-broadening around the initial concentration estimate.) The total air pressure is fixed based on the average of sensor readings at the rooftop and mirror that were calibrated to within ±30 Pa with a recently factory-calibrated manometer. The dual-comb system returns the spectrum measured simultaneously at the two comb teeth separated by some integer multiple of $\Delta f_r$; this effect is included in the simulation to remove even this small instrument line shape, but it was found to be negligible. For the lower spectral region, the least-squares fit is performed in the same manner except that the temperature is fixed at the value returned from the fit across the $CO_2$ band in the upper region.

Over the 170-minute spectrum of Fig. 1d and 2, the variations in average temperature, average pressure, and dry-air volume mole fractions were, respectively: $T$ = 293±0.5 K, $P$ = 82250±25 Pa, $[CO_2]_{VMR}$=397.6 ±0.26 ppm, $[CH_4]_{VMR}$ =1874±0.38 ppb, and $[H_2O]_{VMR}$ = 3168±170 ppm.